\documentclass{article}
\usepackage{graphicx} % Required for inserting images
\usepackage[numbers]{natbib}
\usepackage{listings}
\usepackage{color}
\usepackage{subcaption}
\usepackage{hyperref}

\definecolor{dkgreen}{rgb}{0,0.6,0}
\definecolor{gray}{rgb}{0.5,0.5,0.5}
\definecolor{mauve}{rgb}{0.58,0,0.82}

\newcommand{\footremember}[2]{%
    \footnote{#2}
    \newcounter{#1}
    \setcounter{#1}{\value{footnote}}%
}
\newcommand{\footrecall}[1]{%
    \footnotemark[\value{#1}]%
}

\title{Lomas: A Platform for Confidential Analysis of Private Data}

\author{Damien Aymon\footremember{equal}{Equal contribution.}$^{1}$, Dan-Thuy Lam\footrecall{equal} $^{1}$, \\ Lancelot Marti\footrecall{equal} $^{1}$, Pauline Maury-Laribière\footrecall{equal} $^{1}$, \\  Christine Choirat$^{1,2}$, and Raphaël de Fondeville$^{1}$ \vspace{1em} \\ $^{1}$ Federal Statistical Office, Neuchâtel, Switzerland \\ $^{2}$ University of Geneva, Geneva, Switzerland}

\begin{document}

\maketitle

\begin{abstract}
Public services collect massive volumes of data to fulfill their missions. These data fuel the generation of regional, national, and international statistics across various sectors. However, their immense potential remains largely untapped due to strict and legitimate privacy regulations. In this context, Lomas is a novel open-source platform designed to realize the full potential of the data held by public administrations. It enables authorized users, such as approved researchers and government analysts, to execute algorithms on confidential datasets without directly accessing the data. The Lomas platform is designed to operate within a trusted computing environment, such as governmental IT infrastructure. Authorized users access the platform remotely to submit their algorithms for execution on private datasets. Lomas executes these algorithms without revealing the data to the user and returns the results protected by Differential Privacy, a framework that introduces controlled noise to the results, rendering any attempt to extract identifiable information unreliable. Differential Privacy allows for the mathematical quantification and control of the risk of disclosure while allowing for a complete transparency regarding how data is protected and utilized. The contributions of this project will significantly transform how data held by public services are used, unlocking valuable insights from previously inaccessible data. Lomas empowers research, informing policy development, e.g., public health interventions, and driving innovation across sectors, all while upholding the highest data confidentiality standards.
 %This approach empowers research, advances policy development, e.g., informing public health interventions, and drives innovation across various sectors while upholding the highest standards of data confidentiality.

\end{abstract}

\section{Introduction}\label{sec:intro}

Public services possess a vast amount of data that are indispensable for carrying out their public-interest missions. States invest significant time and resources in gathering, sorting, and consolidating these datasets, which serve as a unique and dependable source of information. However, despite their inherent value, their full utilization for the betterment of public welfare remains incomplete. These data are strictly earmarked for specific purposes, adhering to data protection regulations such as the General Data Protection Framework (GDPR) \citep{gdpr2016} in the European Union (EU) or the Data Protection Act \citep{lpd2020} in Switzerland. 
Public entities such has National Statistical Offices (NSOs) and statistical agencies are constrained by even stricter laws such as Title 13 \citep{title13} for the US Census Bureau or the regulation on statistical confidentiality \citep{statConfEU2009} in the EU.
These offices are indeed required to protect the confidentiality of the data they collect, which is legally different than privacy \citep[p. 197,222]{usaFedStatCom1994}, and requires a higher degree of protection to ensure their secrecy, and cannot use the data for any other purposes than the publication of national statistics.
Exceptions exists, such as for academic research \citep{researchEU2009} in the European Union, for which any request for data sharing undergoes a thorough review to evaluate its legitimacy and the risk of identifiable data disclosure. Typically, the administrative process for accessing such data in these instances is lengthy and intricate.

These oases of data remains unreachable and the full potential of the data collected by public services can only be realized through their "secondary utilization", i.e., using these data for purposes other than those for which they were collected:
\begin{itemize}
\item Other governmental offices can exploit these data to effectively support the public policy-making process based on objective indicators. For instance, a National Statistical Office can furnish accurate demographic and economic data to inform policy decisions regarding public health or education.
\item Research laboratories can employ this data to advance knowledge and enhance a country's competitiveness through innovation. For example, university researchers could analyze data held by governmental agencies to devise innovative solutions to address poverty or school dropout rates.
\item The private sector can exploit these data to eventually bolster national economic vitality. For instance, companies could utilize data collected by public services to motivate strategic decisions regarding investment or product development.
\item These data can also made available to other states to better address global crises such as the COVID-19 epidemic. For instance, a country could use epidemiological data detained by other nations to facilitate coordinating an effective global response to the pandemic.
\end{itemize}
However, it is imperative that this data reuse is politically and socially accepted, ensuring that the benefits it brings do not come at the expense of citizens, households, and businesses. To uphold trust in the state, it is therefore crucial to control the risk of disclosing identifiable data when they are used both within and outside the public sector.
Lomas attempts to answer this societal challenge by offering a service enabling "eyes-off data science", i.e., a practice of data science where private data is never accessed directly by its practitioner. The platform facilitates the realization of the full potential of private data while formally managing the disclosure risk of identifiable data, thereby safeguarding organizational reputation and mitigating adverse consequences for contributing individuals.

The Lomas platform follows the pioneering experiment initiated by the UN Pet Lab hackathon~\citep{unPetLab2022} with the objective of demonstrating that the secondary utilization of data collected by a NGO was possible thanks to Privacy-Enhancing-Technologies (PETs). The hackathon's platform had been developed by the company Oblivious \footnote{https://www.oblivious.com/}, who open sourced the competition's code serving as Lomas' starting point. 
The platform serves as a hub where authorized entities seeking to repurpose data can connect, generating {data products}, i.e., the results of any data processing algorithm such as descriptive statistics or trained machine learning models,
%, with controlled disclosure risks,
while controlling disclosure risks and so safeguarding the original data at the confidential unit level from direct exposure to users.
%This ensures that private data does not circulate between {data owners}, i.e., any public entity such as a NSO, and the {data user}.

Lomas is a "remote access" service, however, it adds upon conventional solutions in two key aspects: Firstly, it guarantees that users never directly access identifiable data
%, implementing the concept of "eyes-off data science"
and secondly, it eliminates the need for human intervention in conducting output checks on the algorithms executed by the user. The latter typically necessitates rigorous scrutiny of algorithmic results by experts in disclosure risk, a process that is both expensive and time-intensive. Employing PETs enables the automation of the entire process while maintaining formal oversight of the risk of disclosure.

Consequently, the platform unlocks the value of data held by public services through their secondary usage, which may have been deemed so far too sensitive or prohibited under the current legislation. 
Leveraging PETs enables precise management of data accesses and disclosure risks regardless of the nature of private data usage, so that the risk of harmful consequences for contributing individuals is controlled when their data is re-used: A sufficient mitigation of this risk contributes to safeguard public trust.
Alternatively, the platform can also be used to expedite the development of data analysis algorithms while awaiting formal data sharing procedures prescribed by law, when they exist. 

To stress the platform's added-value, let us consider the utilization of data for research as an illustrative example. Presently, processes for sharing data from public services are characterized by their slowness and complexity necessitating the establishment of a formal legal agreement and heavy data pre-processing; tasks that both require the direct intervention of public servants at all hierarchical levels. 
%This framework often transfers the responsibility for data protection to the research laboratory.
In addition to its considerable administrative overhead, this procedure markedly diminishes the pace of research—a circumstance not always compatible with international crises such as the Covid-19 pandemic.
For research, Lomas holds significant relevance in the following scenarios:
\begin{enumerate}
\item Facilitation and/or preliminary work: This approach aims to expedite data provision to researchers, enabling them to commence their work promptly; they can then run feasibility assessment of analyses, prepare and test algorithms while awaiting the conclusion of the administrative process for obtaining the original data. The advantages are twofold:
\begin{itemize}
\item For the researcher: Acceleration of the research project.
\item For the public administration: Resource savings in cases where analyses are deemed to be unfeasible.
\end{itemize}
\item Enabling Accessibility: This case involves providing access to data that was previously considered too sensitive, under the condition that these are never disclosed in their original form and the risk of disclosure is meticulously managed. The advantages are outlined as follows:
\begin{itemize}
    \item For the user: Analyze to data that would typically be inaccessible.
\item For the government: Achieving the full potential inherent in its collected data. Moreover, there is a favorable reputational outcome, as the financial commitment invested in gathering and consolidating datasets extends beyond the originating governmental office.
\end{itemize}
\end{enumerate}
%For instance, the platform can serve as an interim solution until a research laboratory gains access to the original data, allowing the latter to start designing algorithms significantly earlier and thus speeding up the research cycle.

To ensure data confidentiality on the platform, algorithmic outputs undergo random perturbation through Differential Privacy (DP) techniques.
For high level of perturbations, access to the platform can be granted with minimal contractual obligations as the disclosure risk can be mathematically proved very low. While the obtained results may exhibit limitations in usefulness, this opportunity remains valuable as it allows for instance researchers to swiftly commence work by testing their code and experimental protocols.

%The platform is a "remote access" service; however, it adds upon conventional solutions in two key aspects. Firstly, it guarantees that users never directly access identifiable data, implementing the concept of "eyes-off data science". Secondly, it eliminates the need for human intervention in conducting output checks on the algorithms executed by the user. The latter typically necessitates rigorous scrutiny of algorithmic results by experts in disclosure risk, a process that is both expensive and time-intensive. Employing DP enables the automation of the entire process while maintaining formal oversight of the risk of disclosure.

%The Lomas platform facilitates the realization of the full potential of private data while formally managing the disclosure risk of identifiable data, thereby safeguarding organizational reputation and mitigating adverse consequences for contributing individuals.

To the best of our knowledge, Lomas is the first open-source platform of its kind. Beyond its innovative design, it stands out as the only platform developed by the public sector for the public sector, targeting public servants and researchers. Although Lomas primarily serves this audience, it also hold the potential to facilitate national and international collaborations with the private sector and NGOs. To ensure broad availability and ease of deployment and maintenance, we have partnered with INSEE, the French NSO, and integrated Lomas into their datalab platform, Onyxia. Through this partnership, we aim to make Lomas widely accessible, lowering barriers to the adoption of PETs and creating a valuable public good.

The article is structured as follows: In Section \ref{sec:pets}, we start by discussing PETs and how trust relationships help decide which technology to use. Section \ref{sec:trust_and_dp} delves deeper into the dynamics of trust relationships for public administrations, concurrently introducing Differential Privacy as a means to mitigate the risk of data disclosure on the platform.  Subsequently, Section \ref{sec:platform} elucidates the foundational design principles guiding Lomas' development, underscoring the pivotal role of metadata for the platform's functionalities and presents the service's features from the user's perspective. %Finally, we present the challenges that such a platform poses and our plans for future developments.
Lastly, we address the challenges inherent in such a platform and outline our plans for future enhancements.

\section{Overview of Privacy-Enhancing Technologies}\label{sec:pets}

Privacy-Enhancing Technologies safeguard data confidentiality throughout the analysis and dissemination of sensitive information. Their utilization allows public services such as NSOs to realize the full benefits of the data they collect while simultaneously minimizing privacy risks for those entrusting their data. Specifically, PETs address the lack of trust among the parties involved in the data value chain, namely
\begin{itemize}
    \item An identifiable person, often referred to as a privacy unit, pertains to any natural or legal entity, or a collection thereof, such as a household, producing information not publicly available and necessitating protection. In the instance of natural persons, an identifiable individual is also called a data subject. Consequently, "identifiable data" encompasses all information gathered regarding an identifiable person. Within the framework of data protection legislations like the Swiss Data Protection Act or the GDPR, which primarily concern natural persons, the term "personal data" is employed.
    \item Input party(ies): An entity responsible for the collection of identifiable data and tasked with its value enhancement through the creation of a data product. Examples include a governmental offices, hospitals, schools and National Statistial Offices which generate statistics such as a poverty rates.
    \item Computing Party(ies): An entity responsible for providing the computational resources essential for developing the data product. Computations commonly occur on remote servers, usually through a cloud infrastructure. This infrastructure may be managed either by a trusted organization, such as a governmental agency, or by a private enterprise offering access via the public Internet to servers utilized possibly by multiple clients, thus often referred as public cloud services.
    \item Result Party(ies): The group of individuals to whom the data product is disseminated. For instance, in the case of a public authority, any Open Government Data product is distributed without constraints to the public, representing the broadest distribution circle where confidentiality controls must be stringent due to the heightened risk of extracting identifiable data. Alternatively, the dissemination of a product may be confined to narrower circles, such as public administration employees, all of whom are bound by official, or statistical, secrecy, thereby reducing the risk of attempting to extract identifiable data compared to distribution to the general public.
\end{itemize}

The trust relationship among the aforementioned parties dictates the level of confidentiality that must be upheld throughout the data value chain. In this context, "trust" encompasses any rationale justifying the non-disclosure of identifiable data to another party. This rationale may include, for instance, legal prohibitions against the exchange of identifiable data between parties, distrust from data subjects toward the data-providing entity, or the lack of legal frameworks or organizational measures to mitigate attempts to extract identifiable data. Overcoming the absence of trust between multiple stakeholders can be achieved through the adoption of privacy-enhancing technologies, which can be broadly categorized into two categories:
    \begin{itemize}
        \item Input Privacy technologies are designed to enable two or more stakeholders described above to contribute data for computation without any other party accessing the original data in clear text, i.e., unencrypted or unprotected. These technologies typically rely on cryptographic methods, or dedicated hardware, and are thus integral components of data security frameworks.
        \item Output Privacy technologies are specifically engineered to restrict the extraction of identifiable data from a data product. It is noteworthy that within this framework, the data product is voluntarily released by the data provider. This procedure is also known as (statistical) disclosure control, a domain well-established in NSO, which have honed this practice over numerous years.
    \end{itemize}
Thus, contingent upon the trust relationships among the diverse stakeholders within the data value chain, the entity responsible for ensuring confidentiality, typically the input party, will use a combination of PETs to safeguard the confidentiality of the inputs and outputs.

% ---------------------------------------------
% Section - General principle and key features
% ---------------------------------------------
\section{Lomas Trust Schema and Differential Privacy}\label{sec:trust_and_dp}
In this section, we start by describing trust relationships in public administrations to motivate the choice of the PETs used in Lomas and explain in detail why Differential Privacy is most natural in this context.

\subsection{Trust Relationships in Public Services}
\begin{figure}
    \centering
\includegraphics[width=0.8\textwidth]{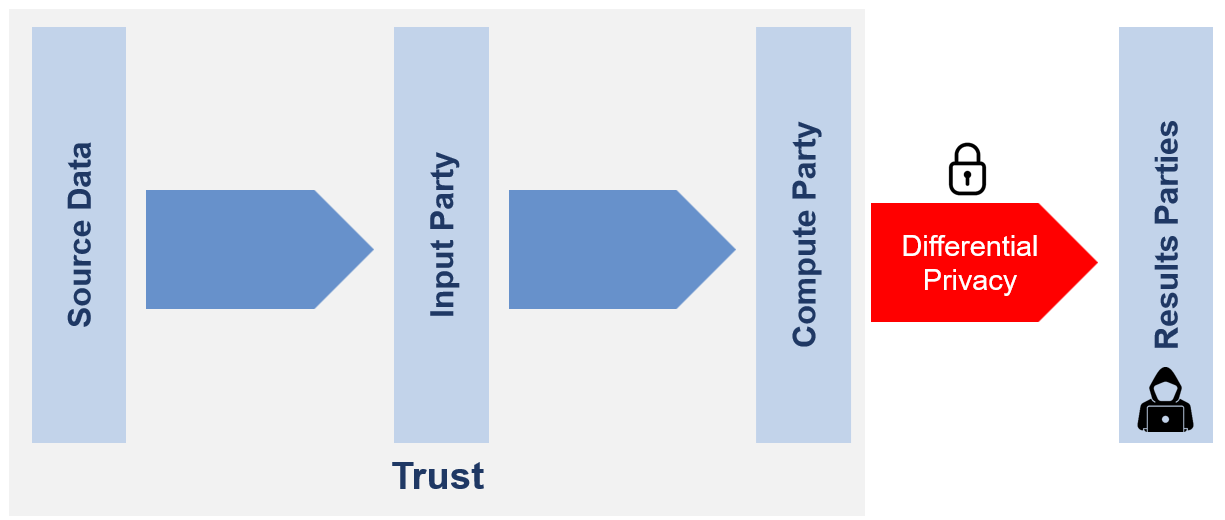}
    \caption{Trust schema for Lomas, a platform for public services to enable analysis of their private data to other parties while controlling the risk of disclosure. This schema is common for public administrations which can rely on trusted IT infrastructures.}
    \label{fig:trust_schema}
\end{figure}

The architecture of the platform is structured in accordance with the trust schema depicted in Figure \ref{fig:trust_schema}, aiming to make extensive use of relationships specificities within public administrations:
\begin{itemize}
    \item  Confidentiality units transmit their data to the input party, typically a governmental office. The trust relationship is commonly enforced through established legal frameworks.
\item Public agencies often have access to a trusted computing provider, either through their own infrastructure or a centralized provider such as a Governmental Office for Information Technology, Systems, and Telecommunication. This arrangement obviates the necessity for input privacy-enhancing-technologies. The provider assumes responsibility for data security, including implementing requisite measures to prevent unauthorized data access on the platform. Additionally, the provider oversees user identification and authentication.
\item The eligible Results Parties may vary, encompassing governmental employees from entities other than the data-collecting office to the general public. In all scenarios, it is presumed that no inherent trust relationship exists between the input party (the office providing the data) and the results party. Consequently, the deployment of output privacy technologies, notably differential privacy, becomes imperative to mitigate the risk of disclosing confidential information.
\end{itemize}
This trust model is prevalent in public administrations, which have access to trusted computing resources directly managed by the state or other certified providers.

\subsection{Disclosure Control Mechanisms}
From a strictly technical standpoint, recent scientific advancements underscore the inability to ensure complete anonymization, i.e., irreversibly eliminate the risk of identifiable data disclosure: the publication of any data product, regardless of its nature, inherently carries a non-zero risk of identifiable data exposure. Nonetheless, this risk can be managed through technical means. These frameworks facilitate the management of what is commonly referred to as the privacy-utility trade-off: essentially, this concept elucidates the challenge of achieving both high levels of confidentiality and data product utility simultaneously. Consequently, significantly mitigating the risk of disclosure often necessitates sacrificing product utility, and vice versa. The output privacy-enhancing technologies enable the quantification and informed management of this trade-off. Conversely, determining an acceptable level of disclosure risk and/or minimum utility is a complex socio-political matter that cannot be solely addressed through technological means.

In the absence of control over auxiliary sources of information available to potential attackers, un-protected data products become susceptible to reconstruction \citep{abowd2023}, linkage \citep{sweeney2000}, and membership \citep{homer2008} attacks.
Through the process of randomizing algorithm outputs, Differential Privacy (DP) \cite{dwork2006} renders any attempts to extract identifiable data from the data product unreliable, thereby mitigating its disclosure risk.
The latter offers several key properties that contribute to its effectiveness in safeguarding data privacy:
    \begin{itemize}
    \item Firstly, DP provides a robust privacy guarantee, ensuring precise control over the risk of identifiable data disclosure. This guarantee remains steadfast regardless of the auxiliary information and technical capabilities available to potential attackers.
    \item Secondly, DP's post-processing property ensures resilience against any form of data transformation. This robustness allows for full transparency regarding the randomization mechanism employed for data protection, thereby enhancing trust and accountability in the privacy-preserving process.
    \item Lastly, DP's composition property enables the quantification of disclosure risks across multiple data products. By tracking risks independently of the nature of the data products, DP facilitates the comprehensive assessment of disclosure risks at the confidential unit level, thus enhancing overall privacy protection.
    \end{itemize}
The quantification of this risk is achieved through a parameter known as the privacy-loss budget, the selection of which is contingent upon both the sensitivity of the data and the intended utilization of the safeguarded data product.

In order for a data analysis to be sanctioned for execution on the Lomas platform, it necessitates the inclusion of a disclosure control mechanism, specifically a randomization mechanism that adheres to the conditions of differential privacy.
The project does not aim to develop these algorithms internally, but rather relies on established open-source libraries such as OpenDP \citep{opendp2020} and SmartNoiseSQL \citep{smartnoiseSQL} for data manipulation and transformation, SmartNoise-Synth \citep{smartnoiseSynth} for generation of syntetic datasets as well as DiffPrivLib \citep{Holohan2019} for machine learning and artificial intelligence tasks.

\section{Platform Design and Features}\label{sec:platform}
This section first outlines the principles guiding the development of Lomas, with the objective of developing the platform as a portable, open-source, community-driven standard product for public services.
Then, we provide a detailed description of the workflow from the user's perspective, illustrating the processes and interactions involved in utilizing Lomas. Details of the Lomas architecture and deployment can be found in Appendix \ref{sec:appendix_a}.
Finally, we emphasize the importance of metadata as a crucial element for the automated application of DP. We dedicate a portion of this document to explain the reasons behind this emphasis and its implications for the effectiveness of Lomas.

\subsection{Design Principles}
We start by presenting the principles underpinning the architectural design of Lomas:
\paragraph{Open-source} Building upon the code developed for the 2022 UN Pet Lab Hackathon, the Lomas platform is open-source and available under a MIT license. The latter gives the rights without limitation to use, copy, modify, merge, publish, distribute, sublicense, and/or sell copies of the Software. The code is available on the GitHub of the Data Science Competence Center \footnote{https://github.com/dscc-admin-ch/lomas} of the Swiss Federal Statistical Office. The project is open to contributions by the community to incorporate new functionalities.

\paragraph{Modular} The objective of the service is not to furnish customized implementations of existing services and/or libraries for which trusted and validated open-source implementations already exist. Instead, it depends on the incorporation of established open-source libraries such as "openDP," "SmartNoise-SQL", SmartNoise-Synth", and "DiffPrivLib." The inclusion of libraries not yet provided by the service can be facilitated either through a proposal to the development team or via direct contribution to the project on GitHub$^{3}$. The service also relies on established data management and storage services for which integrators for most popular solutions, e.g., tabular data files, S3, \dots are, or will be, offered. 

\paragraph{Portable} With the aim of positioning Lomas as a standard platform provided by public services, it has been designed to maximize its portability. Consequently, its components have been either containerized or integrated into a library, enabling deployment or installation on standard infrastructure.

\paragraph{Easy to use} The platform's value proposition revolves around streamlining access to differentially private data science. This simplification manifests in two key aspects:
\begin{itemize}
    \item The deployment of the service, from both the operator's and the user's perspectives, is prioritized to be as straightforward and automated as possible. For instance, we offer Helm charts for the automatic deployment of Lomas' containers; furthermore, the service is available for testing as a "click-and-deploy" service on the data science platform Onyxia \footnote{https://datalab.sspcloud.fr/}, developed by the INSEE, the french NSO.
    \item the usage of libraries implementing differential privacy is simplified as much as possible, for instance by automatically populating parameters obtainable from the metadata catalog.
\end{itemize}

\paragraph{Scalable} The containerized architecture of the service inherently enables the platform to scale alongside the computing resources accessible on the infrastructure utilized for deployment. Regarding scalability of dataset sizes, the service's capability to manage large datasets hinges on the implementation of DP libraries: certain services may necessitate loading the entire dataset into memory, whereas others facilitate streaming or distributed computations. The modular design of Lomas is pivotal in ensuring the scalability of the service.

\paragraph{Secure} From a security standpoint, the Lomas platform is not yet production-ready and currently lacks the implementation of security protocols for authentication and connections. These aspects will require further adaptations during a potential deployment in production. Nonetheless, Lomas employs input query verification using Pydantic \citep{pydantic2024} and entrusts the evaluation of DP-protected pipelines to the respective libraries. Additionally, certain measures to mitigate timing attacks have been implemented.

\paragraph{Client-server architecture} The Lomas platform is structured around two primary components: a client library and an HTTP server.
The former offers a set of commands allowing an accredited user to send DP-protected algorithms to the server, while the latter serves as a gate-keeper between the user and the private dataset by processing the client's requests and managing an administration database storing for instance user identities, budgets, and DP-algorithms archives. 
The organization aiming to offer the service is responsible for its deployment as well as managing users and available datasets by adding, modifying or deleting information in the administration database.
Following the modularity principle, the Lomas service is not intended as a substitute for existing data storage infrastructure. Instead, it relies on pre-existing services to store and administer private datasets. The service's purpose is solely to furnish the requisite tools for computing differentially private algorithms on designated datasets, while managing their access paths in the administration database along with their metadata.
%Thus, we implement adapters allowing the service to connect to various kinds of remote databases and fetch the sensitive data to be queried on.

%This section first describes in detail the client library's functionalities and step-by-step workflow from the user point of view. We then present in detail server implementations.

\subsection{Lomas Service's Description}\label{sec:lomas_service}
\begin{figure}
    \centering
\includegraphics[width=0.9\textwidth]{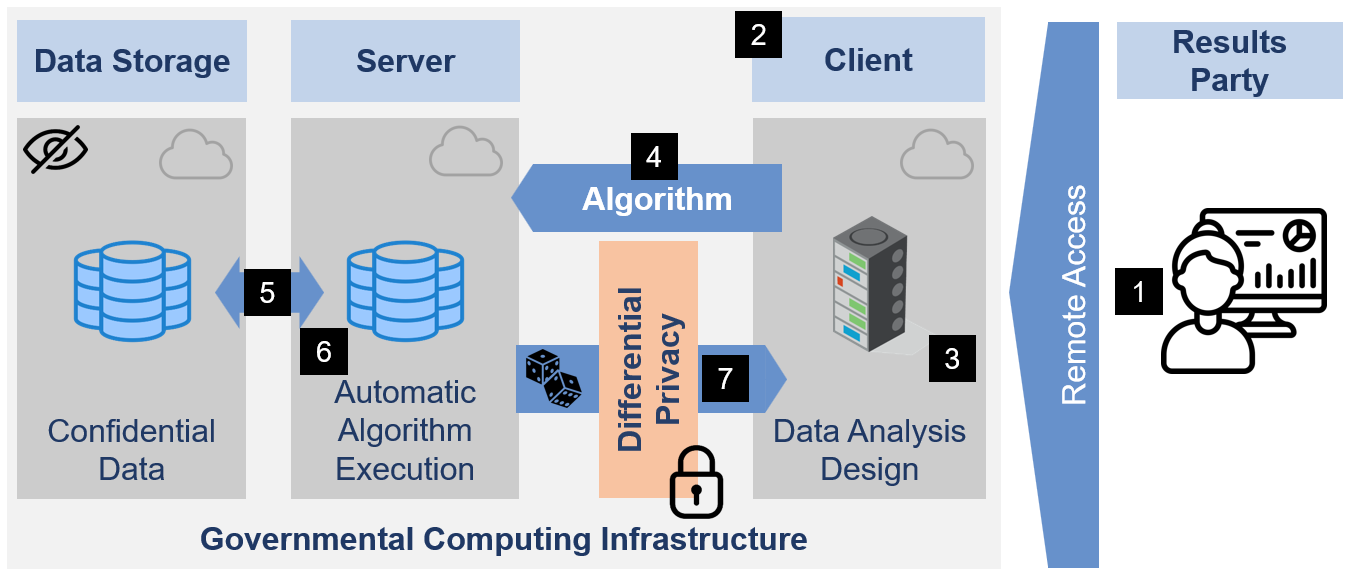}
    \caption{Schematic diagram of Lomas, a platform for public services to enable analysis of their own private data to other parties while controlling the risk of disclosure. Numbers in black squares refer to the user's step-by-step workflow whose description can be found in Section \ref{sec:lomas_service}.}
    \label{fig:service_schema}
\end{figure}

The service's step-by-step workflow is depicted in Figure \ref{fig:service_schema}, elucidating the process by which data is analyzed by authorized users. From the user's perspective, the service operates along the following steps that matches the numbers in Figure \ref{fig:service_schema}:
\begin{enumerate}
    \item The user establishes a remote connection to the platform, gaining access to a virtual environment tailored for executing Python programming language scripts.
    %Proficiency in this programming language is assumed as it is prevalent in data analysis across private and academic sectors.
    Extending compatibility to the R programming environment will be considered in future developments.
\item Utilizing the "\verb|lomas-client|" library, users can review a catalog of datasets that have been made available to them. This catalog includes comprehensive information regarding the nature of the datasets, their representation (metadata), and the currently allocated privacy-loss budget.
\item Upon selecting a dataset, users can download a fully simulated dummy dataset into their computational environment, faithfully replicating the metadata attributes of the original dataset. This dummy dataset serves as a means to design and test data analysis algorithms without spending their privacy-loss budget.
\item Subsequently, users transmit their finalized algorithms to a remote server managed by the public administration. Prior to execution, the algorithm undergoes scrutiny to ensure the presence of a confidentiality control mechanism, specifically a differentially private mechanism safeguarding the algorithm's output. If such a mechanism is absent, the algorithm's execution is declined. Additionally, the server verifies the user's dataset access rights and the adequacy of their privacy-loss budget.
\item	If the access and confidentiality requirements are met, the secure server retrieves the private data from the secure storage space.
\item	The algorithm is autonomously applied to the original dataset, with the resultant calculation being protected via differential privacy.
\item	The protected result is saved and transmitted back to the user. Logging algorithms and their DP-protected results enables authorities to monitor user activity and datasets' utilization.
\end{enumerate}
This approach empowers users to execute algorithms without direct access to private data, while simultaneously assuring the public agency providing the data that the risk of disclosing confidential information is minimized.

\subsection{Metadata Catalog}

Metadata pertains to descriptive information concerning data, commonly referred to as 'data about data'. It furnishes supplementary details that characterize a piece of information, distinct from the information's content itself. For instance, within the context of a dataset table, the data refers to the actual cell values, whereas metadata encompasses attributes such as column names and types.
In the pursuit of "eyes-off data science", as enabled by Lomas, metadata bears critical importance:
\begin{itemize}
    \item Users depend on metadata to comprehend the available data and request allocation of a privacy-loss budget.
    \item Metadata facilitates the service's capacity to dynamically generate realistic dummy datasets, enabling the design and testing of algorithms without depleting the privacy-loss budget.
    \item The service relies on metadata for the autonomous execution of differentially private pipelines without requiring user intervention. For instance, knowledge of the range of possible values for an attribute is indispensable for applying the appropriate level of noise to ensure the desired privacy guarantee.
\end{itemize}
Due to these considerations, metadata in Lomas are assumed to be publicly accessible, as is typically the case for data held by public services, or at least freely available to the platform's users.

With the increasing adoption by public administrations of the "once only principle" \citep{kush2020, oop2016,oop2022}, i.e., that data from citizens and businesses are collected by public services only once, the platform benefits from initiatives aimed at cataloging and harmonizing the datasets maintained by public entities, particularly focusing on documenting their metadata. Interoperability standards, such as the DCAT Application Profile for data portals in Europe \citep{dcatap2024} and data catalogs such as the I14Y platform in Switzerland %\citep{i14y2023}
\footnote{https://www.i14y.admin.ch/en/home}
, simplify the generation of dummy datasets for assessing user-designed data analysis algorithms. The direct integration of these standards into Lomas would enable the platform to leverage these initiatives synergistically, thereby fully exploiting the value of state-held data.

% ---------------------------------------------
% Section - Challenges and future developments
% ---------------------------------------------
\section{Challenges and Future Developments}\label{sec:challenges}

Lomas capitalizes on the distinctive attributes inherent in trust relationships within public administrations. Through collaboration with a trusted computing party, the platform's focus is placed only on ensuring the confidentiality of the generated results. This objective is realized through the implementation of differential privacy, whose primary challenge pertains to the determination of the privacy-loss budget. Indeed, the DP framework enables precise control over the balance between data utility and confidentiality by allocating a maximum privacy-loss budget for each user. The selection of this budget entails a consideration that has political implications and can be articulated as follows: "What is the maximal acceptable level of risk to which the dataset's confidentiality units can be exposed?" or "What is the minimal intended purpose for which the data product must reliably be used?".
These inquiries align with the "confidentiality first" and "utility first" principles, respectively, allowing an organization to delineate its priorities and judiciously determine the privacy-loss budget through informed decision-making.

To illustrate this compromise and decision-making process, let us consider again the utilization of data in research as an illustrative example.
The two scenarios described in Section \ref{sec:intro}, namely "Facilitation and/or preliminary work" and "Enabling Accessibility", both call for a small privacy-loss budget, i.e., large random perturbations, to compensate for the high sensitivity of the project, whether due to the sensitivity of the data or the short time frame:
%, which makes it difficult to set up extensive controls on potential users or to conclude formal contracts.
The high risk yielded by either the sensitive nature of the data, or the urgency of the request, is mitigated thanks to an adequate privacy-loss budget policy that reflects the context in which the data value enhancement takes place.

Lomas thus hinges upon the establishment of a policy governing the authorizations of data product's creation and the selection of an adequate privacy-loss budget. 
Such policy cannot be comprehensively prescribed solely through a technological lens; it must consider the broader ecosystem, including the nature of the project, sensitivity of the data, the circle of individuals accessing the data product, the computing environment, the existing legal framework, and, notably, the objective assessment of the resultant disclosure risk. 
Its establishment thus necessitates a data governance framework such as the "Five Safes" \citep{desai2016} encompassing all dimensions of the data value enhancement process, namely Project, Data, People, Setting, and Output, to mediate discussions among privacy professionals, decision-makers, and citizens.
Confidentiality alone does not ensure social acceptance of data secondary usage: the project's alignment with the rule of law and its public benefits also comes into play and for those, an objective quantification using a technological solution is not possible.

There are very active ongoing efforts to help bridge the gap between technology and policy. The recent US Executive Order on Artificial Intelligence \citep{eo2023} emphasizes the importance of Privacy-Enhancing Technologies (PETs) like differential privacy – the core technology behind Lomas. This Executive Order requests federal agencies to integrate AI technologies in a manner that ensures the protection of individual privacy and national security. To this end, the National Institute of Standards and Technology (NIST) has been tasked with developing standards, guidelines
and certification mechanisms, which aim to simplify the implementation of differential privacy \citep{nist2023}.
These initiatives aim to eliminate the need for users to grasp the intricacies of differential privacy guarantees, thus fostering broader adoption and trust in these frameworks.

Currently, Lomas integrates open-source libraries that implement differential privacy for SQL queries and classical data analytics. We plan to extend its functionalities to support the training of machine learning algorithms and the creation of synthetic datasets. Thanks to the metadata catalog, which we plan to align on international standards by extending frameworks such as DCAT to meet differential privacy requirements, future developments will include improved automatic parameters filling in differential privacy algorithms, improving and simplifying the platform's usability. Additionally, from a system engineering perspective, the platform will implement standard security protocols to ensure secure connections between its components, manage users authentication and broaden the range of dataset storage types handled by the service.

\section*{Acknowledgment}
The Lomas development team would like to sincerely thank the company Oblivious for sharing and open-sourcing the code behind the 2022 UN Pet Lab's Hackathon and the Onyxia development team at INSEE for their support in deploying Lomas on their platform.

%\section*{References}
\bibliography{lomas}
\bibliographystyle{plain}

\appendix

\section{Detailed Service's Architecture and Deployment}\label{sec:appendix_a}

\subsection{Client Library: Functionalities and Workflow}\label{sec:client}

% intro to client library
Submitting requests to the server is done through the Python library \verb|lomas-client|\footnote{https://pypi.org/project/lomas-client/}, which can be installed using \verb|pip install lomas-client|. This ensures synchronization of library versions between the client and server, a crucial requirement for the platform's seamless operation. The library provides the following functionalities:
\begin{itemize}
    \item Algorithm Serialization: the library transforms DP-protected algorithms into JSON objects that can be de-serialized by the server. The serialization is tailored for each available DP-library.
    %the various methods prepare the body of their server request in JSON format. Simple server function parameters can be included as-is in the JSON request body. For example, these include SQL queries to be executed with the SmartnoiseSQL library. However, Python objects such as "opendp" pipelines must be transformed to an intermediate representation that can be included in the JSON request body. To achieve this, the "opendp-logger" package is utilized to convert the "opendp" pipeline object into a text representation in JSON format. The "opendp-logger" package limits its scope to serializing and deserializing "opendp" pipelines and as such, reduces the security risk involved with deserialization of arbitrary objects.
    %In most cases, the parameter values need only to be stored in a JSON object without any further processing. However, the process becomes more involved when dealing with complex objects such as "opendp" pipelines, which are "opendp"-specific objects and must be serialized in order to be included in the request body. To achieve this, the "opendp-logger" library is utilized to convert the "opendp" pipeline object into a serializable JSON object. 
    %Once the DP-algorithm and its parameters are in a serialized format, the request is then finalized: the resulting to the query are stored in the request body, while the user's information are placed in the header.
    \item Communication with the server: serialized DP-algorithms and corresponding DP-protected results are transmitted between the server and the client through a single REST API call.
    \item De-serialization: the library checks the validity of the server's response and, in case of validity, proceeds to its de-serialization in order to deliver the DP-algorithm's result to the user. 
    %Conversely, if the response is deemed invalid, an error message elucidating the issue is returned to the user.
\end{itemize}

We now outline the functionalities of the library, depicting the step-by-step process followed by a user who possesses the authorization to query a sensitive dataset containing personal information about penguins inhabiting diverse islands. The user's goal is to compute the average bill length of the penguin population whose data has been collected. A code example demonstrating all these steps is provided in Figure \ref{fig:code-example}.
%We assume the user is using an environment for data analysis such as JupyterLab, can send API calls to the server and has installed the \verb|lomas-client| Python package.

% login
\paragraph{Client object creation}
%Firstly, the user imports the client's library \verb|lomas-client|. 
The user begins by instantiating a client object, utilizing the server's URL, along with the user's and dataset's names. This client object oversees the requests dispatched to the server and necessitates invocation each time the user intends to submit an algorithm for remote execution.
%will enable her to query the specific dataset using her private budget and access rights.
%The user name and dataset name will be present in the header of all subsequent API requests. 
%Therefore, whenever the user intends to send queries to the server, she needs to call a method on the client object that has been created.
%In the following example and for the rest of this section, the object is named \verb|client|.

\begin{figure}
    \centering
   \begin{lstlisting}[language=Python]
from lomas_client import Client
import polars as pl

# Client object creation
client = Client(
    url = "https://lomas_server.ch",
    user_name = "Dr. Antartica",
    dataset_name = "PENGUIN"
)

# Available privacy-loss budget retrieval
my_initial_buget = client.get_initial_budget() 
my_total_spent_buget = client.get_total_spent_budget()
my_remaining_budget = client.get_remaining_budget()

# Metadata retrieval 
penguin_metadata = client.get_metadata()

# Dummy dataset generation 
dummy_df = client.get_dummy_dataset(seed = 42, nb_rows=100)

# Algorithm design (SQL query using polars)
QUERY = "SELECT AVG(bill_length) FROM df"
with pl.SQLContext(df=pl.from_pandas(dummy_df)) as ctx:
    local_bill_length = ctx.execute(QUERY, eager=True)

# DP-Algorithm Execution Verification on Server
## Remote execution verification
server_bill_length = client.smartnoise_query(
    query = QUERY, epsilon = 100, delta = 0.99, dummy = True, seed = 42
)

## Remote execution consistency check
assert abs(local_bill_length - server_bill_length) < 0.01

# Estimation of privacy-loss budget expenditure
client.estimate_smartnoise_cost(
    query = QUERY, epsilon = 0.1, delta = 0.00001
)

# Query execution on private dataset
avg_bill_length = client.smartnoise_query(
    query = QUERY, epsilon = 0.1, delta = 0.00001
)

# Archives consultation
previous_queries = client.get_previous_queries()

\end{lstlisting}
    \caption{Example of Python code using Lomas to compute the differential private average bill size of a population of penguins.}
    \label{fig:code-example}
\end{figure}

% check budget
\paragraph{Available privacy-loss budget retrieval}
Before sending any algorithm for execution, users can determine the available privacy-loss budget for the dataset being analyzed through the following functionalities:
\begin{itemize}
    \item \verb|get_initial_budget()| retrieves the initial privacy-loss budget that was allocated to the user by the administrator.
    \item \verb|get_total_spent_budget()| provides the total budget spent at the time.
    \item \verb|get_remaining_budget()| returns the remaining available privacy-loss budget. %It is the difference between the initial budget and the total spent budget.
\end{itemize}
Each of these functionalities provides values for both privacy-loss parameters $\epsilon$ and $\delta$.
%If no queries have been conducted yet, the total spent budget should display zero for both $\epsilon$ and $\delta$, while the remaining budget should be equal to the initial budget.

\begin{figure}[h]
    \begin{subfigure}[h]{0.28\textwidth}
        \centering
        \begin{tabular}{|l|l|}
        \hline
        \textbf{island} & \textbf{bill\_length} \\ \hline
        A           & 55.1                  \\ \hline
        B           & 46.1                  \\ \hline
        A           & 50.7                  \\ \hline
        A           & 35.7                  \\ \hline
        B           & 47.0                  \\ \hline
        B           & 51.5                  \\ \hline
        \end{tabular}
       \caption{Private dataset}
       \label{tab:private_dataset}
    \end{subfigure}
    \hfill
    \begin{subfigure}[h]{0.325\textwidth}
        \centering
        \includegraphics[width=1.0\linewidth]{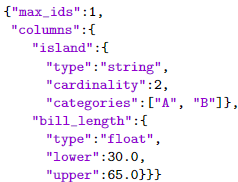} 
        \caption{Metadata}
        \label{tab:metadata}
    \end{subfigure}
    \hfill
    \begin{subfigure}[h]{0.28\textwidth}
        \centering
        \begin{tabular}{|l|l|}
        \hline
        \textbf{island} & \textbf{bill\_length} \\ \hline
        B           & 34.8                  \\ \hline
        A           & 36.4                  \\ \hline
        B           & 34.8                  \\ \hline
        B           & 60.8                  \\ \hline
        B           & 61.2                  \\ \hline
        A           & 39.3                  \\ \hline
        \end{tabular}
        \caption{Dummy dataset}
        \label{tab:dummy_dataset}
     \end{subfigure}
     \caption{Example of private data, metadata, and dummy data for the analysis of the penguin population.}
     \label{fig:data-example}
\end{figure}

% metadata
\paragraph{Metadata retrieval}
To attain a thorough comprehension of the dataset and its contents, users can retrieve the dataset's metadata from the catalog utilizing the \verb|get_dataset_metadata()| function. In Lomas, each dataset is accompanied by a metadata file containing crucial information for the appropriate utilization of Differential Privacy. This includes details such as the maximum number of contributions by privacy unit, column types, and the range of potential values assumed by the variables. An illustration of the function output is provided in Figure \ref{fig:data-example}.
%In our extended version, the metadata can also encompass information such as the number and names of categories in categorical columns.

%For instance, the \verb|get_dataset_metadata()| function details that the table encompasses two columns, one categorical named "island" with two categories ("A" and "B") and one numerical with floating values named "bill length (mm)" with values between 30.0 and 65.0. She also learns that each penguin only appears in one row of the table.

% dummy dataset generation
\paragraph{Dummy dataset generation}
%The metadata contains useful information about the dataset, however, the actual data is stored in tabular form.
For a more concrete comprehension of the dataset's structure, as well as for designing and testing algorithms, users may request the generation of a dummy dataset using the \verb|get_dummy_dataset(nb_rows, seed)| function. 
The latter produces a randomly generated dataset satisfying the constraints imposed by the metadata. The user is allowed to specify a seed for random number generation and the desired number of rows.The dummy dataset, an example of which is provided in Figure \ref{fig:data-example}, serves as a means to locally design and test algorithms before attempting their remote execution on the server.
%The "dummy" dataset is created based on all the information available in the metadata, thus replicating the number, types and name of columns and, if the information is available, numerical bounds and categories.
%With the previous metadata example, it will generate a random DataFrame with two columns: "island" with random sampling between the categories "A" and "B", and "bill length" with random sampling of floating values between 30.0 and 65.0.
%It is noteworthy to mention that this does not give rise to any concerns regarding privacy, as all the information that can be deduced from the fabricated "dummy" dataset is derived from pre-existing, and assumed publicly available, metadata.

% query on dummy - check syntax works
\paragraph{Algorithm design}
%Thanks to the previous steps, the user has gained a comprehensive understanding of the dataset and now seeks to formulate a query to be sent to the server
The user is now prepared to start the design of the algorithm intended for execution on the private data. To prevent unintentional expenditure of the privacy-loss budget at this stage, it is strongly advised to design and test the algorithm locally on the dummy dataset using conventional libraries for data manipulation. For example, \verb|polars| facilitates the testing of SQL queries or data analysis pipelines on the local dummy dataframe. Its compatibility with existing libraries implementing Differential Privacy minimizes the effort required for integrating a differentially private mechanism.

\paragraph{DP-Algorithm execution verification on server}
The user can now proceed to test the execution of the algorithm on the server. To accomplish this, the algorithm must first be modified to incorporate a differentially private mechanism. This can be achieved either by utilizing a DP-library, such as SmartnoiseSQL with \verb|smartnoise_query(query, epsilon, delta, dummy, seed)|, or by integrating a randomization mechanism, for instance, using OpenDP.
Firstly, the user may validate the DP-algorithm by requesting the server to execute it on the "dummy" dataset. This approach allows the user to verify the syntax of the algorithm without incurring any budgetary expenditure. 
%For instance, in order to query the average bill length in the data using "smartnoise-sql", the user can use the function \verb|smartnoise_query(query, epsilon, delta, dummy, seed)|.
%To ensure the validity and effectiveness of the query on the server, the user has the option to set a designated flag called "dummy" which compels the server to execute the query on a randomized "dummy" dataset instead of the actual sensitive dataset. Of course, this operation incurs no expenses as the analysis is conducted on random data. If the user receives a response to her query, it serves as an indication of the correct syntax and validates that if she subsequently submit the same query to the server without the "dummy" flag, she will also acquire a response applied to the real dataset.
%To ensure a query is valid and processed by the server, it is imperative to include both deterministic calculation functions (such as mean and sum) and instructions for adding noise. While "smartnoise-sql" performs this noise addition process automatically, users using "opendp" must remember to  add a random mechanism in their pipeline. Failing to do so will result in a server error, and an error response explaining the issue will be transmitted to the user. As a precautionary measure, it is wise to test and validate queries on dummy datasets before executing them on the actual dataset to avoid any issues.

After confirming the correctness of the syntax, users can validate the algorithm's result on the server by ensuring consistency between the results of remote execution and those obtained from the local non-DP algorithm. As dummy datasets do not consume budget during the execution of DP-algorithms, users can select privacy parameters resulting in negligible noise. This enables users to compare the results of both non-protected and DP-protected algorithms, thereby evaluating the accuracy of remote execution based on the proximity of their outcomes.

\paragraph{Estimation of privacy-loss budget expenditure}
The privacy-loss parameters utilized for the execution of a DP-algorithm may not directly correspond to the actual consumed budget in practice; typically, it tends to be higher contingent upon the algorithm's complexity. Consequently, it is crucial to estimate the expenditure of the privacy-loss budget prior to dispatching the algorithm for processing on the server. For instance, the functions \verb|estimate_smartnoise_cost()| and \verb|estimate_opendp_cost()| provide the accurate budget expenditure for a given request. Accordingly, users have the flexibility to experiment with various values of privacy parameters until they identify those aligning with the privacy-loss budget they are prepared to allocate. It should be emphasized once again that any request submitted to the server for testing on the dummy dataset does not deplete the user's available privacy-loss budget.
%Similarly in "opendp", a specific mechanism's scale is often provided as a parameter, which is not explicitly stated in the $\epsilon$ and $\delta$ budget values format.
%that would be expended by "smartnoise-sql" and "opendp" in terms of $\epsilon$ and $\delta$ respectively.
%It is important to emphasize that the estimation of query cost solely relies on the metadata available to the server, and thus, no actual budget is consumed during this step.

\paragraph{Query execution on original dataset}
Once the privacy parameters have been chosen and the algorithm's execution has been rigorously validated, the user can proceed to apply the algorithm to the private data by simply omitting the \verb|dummy = True| parameter. For example, using \verb|smartnoise_query(query, epsilon, delta)| for "smartnoise-sql" and \verb|opendp_query(opendp_pipeline)| for "opendp" will yield the DP-protected result on the private dataset.
%This can be achieved by utilizing the same function used for querying the "dummy" dataset, albeit without the "dummy" flag.
%It is important to note that only queries performed on the sensitive dataset will consume budget, while the remaining functionalities outlined in the preceding sections will not have any impact on the budget.

\paragraph{Archives consultation}
All executed requests that deplete the privacy-loss budget are archived for subsequent reference by both the user and the platform's administrator. This functionality facilitates the retrieval of past DP-algorithms and their corresponding responses, thereby mitigating the risk of duplicate queries that could exhaust the privacy-loss budget. \\

Additional information regarding the library's functions can be accessed in the \href{https://dscc-admin-ch.github.io/lomas-client-docs/}{GitHub documentation pages}, alongside \href{https://github.com/dscc-admin-ch/lomas/tree/master/client/notebooks}{Jupyter Notebooks} providing further detailed examples. Presently, the client is exclusively accessible in Python. However, it could potentially be ported to alternative programming languages such as R, with the primary constraint being the availability of libraries implementing differential privacy in the respective language.%particularly for applications where a client needs to write pipelines such as "opendp".

\subsection{Server: Components and Functionalities}

% {\color{blue}
% \begin{itemize}
%     \item microservice architecture:
%     \item \begin{itemize}
%         \item fastapi server for request handling and query execution
% 		\item mongodb database for server state (user collection with associated datasets and budgets, references to datasets and dataset metadata, etc.)
% 		\item databases for datasets handled externally to XXX service, currently support adapters to s3, http file download, local files)
%     \end{itemize}
% 	\item fastapi server implementation details:
%     \item \begin{itemize}
%         \item very modular implementation
% 		\item describe basic features implementation (metadata + dummy dataset generation)
% 		\item describe operations that take place for a typical request (i.e. pipeline on dummy/real dataset + budget estimation is same but without returning result + result query is stored and can be recovered later)
%     \end{itemize}
% \end{itemize}
% }

The server comprises two principal components:
\begin{itemize}
    \item A client-facing HTTP server responsible for processing and validating user requests, as well as executing diverse algorithms on the private and dummy data. Its primary function is to efficiently manage incoming requests from clients and execute authorized DP-algorithms.

    \item An administration database, typically a MongoDB by default, which serves as a repository for user and dataset information. User-related data includes access permissions to specific datasets, allocated privacy-loss budgets, remaining budgets, and archives of executed DP-algorithms. Dataset-related data encompasses details such as dataset names, information and credentials for dataset access, along with corresponding metadata.

\end{itemize}

As previously stated, dataset storage is not managed by Lomas. Instead, the platform interfaces with external databases, typically operated by a data provider, to retrieve datasets prior to query execution. Presently, the service includes adapters for S3, HTTP files server, and files stored on locally mounted volumes. 

%\subsubsection{Processing of the user's queries }
Interaction with the client library is brought by a FastAPI application through various HTTP Rest API calls. The functions outlined below furnish the means to administer the server and automate the execution of DP-algorithms:

\paragraph{Service management}
Lomas provides a command line tool for managing the administration database which enables seamless addition, modification, and deletion of users, query archives, and dataset metadata either individually or in batches, provided the correct security credentials are provided.
An illustration of these functions can be found in Figure \ref{fig:server-code-example}

\begin{figure}
    \centering
   \begin{lstlisting}[language=Bash]
# Service deployment in a Kubernetes cluster
# -----------------------------------------------------------------------

# Download Helm chart
helm repo add lomas https://dscc-admin-ch.github.io/helm-charts

# Adapt values.yaml file and run the following to deploy the chart
helm install -f values.yaml lomas-server lomas/lomas-server

# Service administration through CLI tool
# -----------------------------------------------------------------------

# Install script and requirements
git clone https://github.com/dscc-admin-ch/lomas.git
python -m venv venv
./venv/bin/activate
cd ./lomas/server
pip install -r requirements.txt
cd ./src

# Add a user
python mongodb_admin.py --add_user_with_budget --user "Dr. Antartica"\
--dataset PENGUIN --epsilon 10 --delta 0.005

# Add a dataset
python mongodb_admin.py --add_dataset --dataset PENGUIN \
--database_type PATH_DB --dataset_path \
"https://raw.githubusercontent.com/mwaskom/seaborn-data/master/penguins.csv"\
--metadata_database_type PATH_DB \
--metadata_path ../data/collections/metadata/penguin_metadata.yaml

# Check everything was correctly configured
python mongodg_admin.py --show_collection users
python mongodg_admin.py --show_collection datasets
python mongodg_admin.py --show_collection metadata
\end{lstlisting}
    \caption{Example of code to manage Lomas' services.}
    \label{fig:server-code-example}
\end{figure}

\paragraph{Sensitive data requests}
The server handles a typical request from a user, wherein the user requests the execution of a DP-algorithm on a private dataset, as follow:

\begin{enumerate}
    \item The user initiates a request containing the desired DP-algorithm via the client library.

    \item The server validates the user's access to the dataset and verifies the absence of any ongoing algorithms. Should an algorithm already be in progress, the query is not processed, and an error message is returned to the user. 

    \item Subsequently, the DP-algorithm is extracted and reconstructed from the user's request.
    %In the case of "opendp", the pipeline object has been serialized to a textual JSON reprensentation in the client to be sent as part of the request. Thus, we use the library "opendp-logger" to deserialize and reconstruct the opendp pipeline object.
    
    \item A validation process is initiated to assess the compatibility of the algorithm with the corresponding Differential Privacy library. The procedure involves verifying that the necessary protection mechanisms prescribed by the DP library are satisfied. For example, in the case of the OpenDP library, the algorithm must incorporate a measurement indicating the inclusion of a randomization mechanism to safeguard the algorithm's result. If the pipeline fails to pass the validation process, an error message is returned to the user.
    
    \item The privacy-loss budget of the algorithm is computed and compared to the user's remaining budget.
    
    \item If the remaining budget permits, the algorithm is executed and the expended budget is subtracted from the user’s allowance.
   
    \item The algorithm and its associated result are stored in the administration database, facilitating subsequent retrieval, thereby allowing users to re-access their findings as necessary.
    
    \item Upon successful validation of the preceding checks and updating of the budget in the administrative database, the result is transmitted to the user.
\end{enumerate}

\paragraph{Dummy data requests}
The user can apply DP-algorithms on dummy datasets for testing purposes without incurring budgetary expenditures.
The procedure follows the subsequent steps:
\begin{enumerate}
    \item The server retrieves the metadata associated with the dataset stored in the administrative database. Subsequently, the server generates a dataset randomly following the metadata structure (including column types, bounds, names, etc.).
    \item The server reconstructs the DP-algorithm and verifies its validity, employing a process similar to algorithms applied on private data.
    \item The server executes the differentially private algorithm on the dummy data, and the output is transmitted directly to the user through the client library.
\end{enumerate}
For algorithms applied to dummy datasets, the checks and updates pertaining to privacy-loss budget are not conducted, and the algorithms and their results are not archived.

\paragraph{Privacy-loss budget request}
As presented in Section \ref{sec:client}, the server implements functions to compute the budget expenditure of their DP-algorithm. The implementation directly leverages the tools provided by the integrated DP libraries.
%We decided to use tools already existing and developed within the external DP libraries to ensure that the estimated spent budget is exactly what will be spent. % For "smartnoise-SQL", the built-in function $get\_privacy\_cost()$ is used. For "opendp" queries, the cost is estimated with the $map()$ operator on the reconstructed pipeline along while using relevant metadata information.

\paragraph{Administration database requests}
A collection of functions provides the users with the relevant data stored in the administration database, encompassing metadata, archives, initial, and expended budget details.
Therefore, upon reception of such a request along with the associated parameters, the server queries the administration database, post-processes the result if required, and transmits the relevant information back to the client.

\end{document}